%% file: l4dc2025-sample.tex
\documentclass{l4dc2025}

\title[CIKANs]{CIKAN: Constraint Informed Kolmogorov-Arnold Networks for Autonomous Spacecraft Rendezvous using Time Shift Governor}
\usepackage{times}

\SetKwRepeat{Do}{do}{while}%
\usepackage{algorithm}
\usepackage{algpseudocode}
\usepackage{multirow}
\usepackage[export]{adjustbox}

\newtheorem*{assumption*}{\assumptionnumber}
\providecommand{\assumptionnumber}{}


\author{%
 \Name{Taehyeun Kim} \Email{taehyeun@umich.edu}\\
 \Name{Anouck Girard} \Email{anouck@umich.edu}\\
 \Name{Ilya Kolmanovsky} \Email{ilya@umich.edu}\\
 \addr University of Michigan, Ann Arbor, MI, USA
}

\begin{document}

\maketitle

\begin{abstract}%
The paper considers a Constrained-Informed Neural Network (CINN) approximation for the Time Shift Governor (TSG), which is an add-on scheme to the nominal closed-loop system used to enforce constraints by time-shifting the reference trajectory in spacecraft rendezvous applications. We incorporate Kolmogorov-Arnold Networks (KANs), an emerging architecture in the AI community, as a fundamental component of CINN and propose a Constrained-Informed Kolmogorov-Arnold Network (CIKAN)-based approximation for TSG. We demonstrate the effectiveness of the CIKAN-based TSG through simulations of constrained spacecraft rendezvous missions on highly elliptic orbits and present comparisons between CIKANs, MLP-based CINNs, and the conventional TSG.

\end{abstract}


\section{Introduction} \label{sec:intro}
\input{./sections/intro.tex}

\section{Time Shift Governor} \label{sec:tsg}
\input{./sections/tsg.tex}

\section{Constraint-informed Neural Network} \label{sec:cinn}
\input{./sections/cinn.tex}

\section{Constrained Spacecraft Rendezvous and Proximity Operations} \label{sec:rpo}
\input{./sections/spacecraft.tex}

\section{Conclusion} \label{sec:conclusion}
\input{./sections/conclusion.tex}

\acks{}

\bibliography{references}

\end{document}

%% file: sections/intro.tex
Integration of machine learning schemes with control algorithms is a growing research area in the field of control theory. The use of Neural Networks (NNs) based on the Multi-Layer Perceptrons (MLPs) is well established to approximate explicitly model predictive control solutions in various applications, such as temperature management~\citep{drgovna2018approximate} and quadrotor~\citep{chen2022large}. Notably, the use of NNs introduces approximation errors, potentially leading to performance degradation compared to the solution of the underlying optimization problem onboard.

\begin{figure}[htbp!] 
    \centering    
    \includegraphics[width=1.05\linewidth]{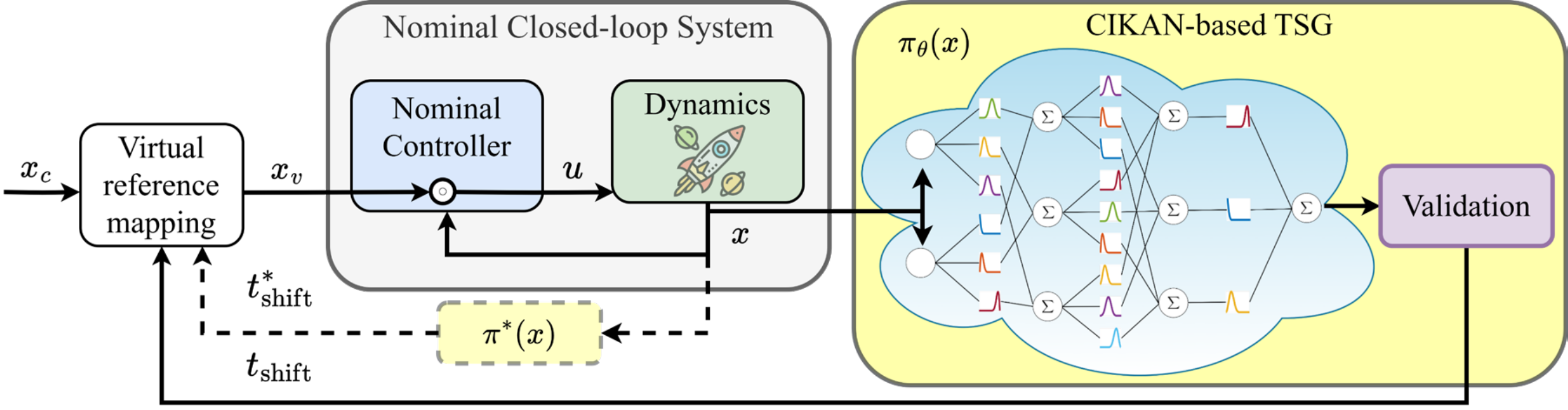}
  \caption{Reference trajectory tracking using a CIKAN-based model: TSG employs the output of the CIKAN-based model to guide the system 
  The nominal controller stabilizes to the virtual reference, $X_{v}$, which guides the system state, $X_d$, }  
  \label{fig:blockDiagram}\vspace{-0.3 in}
\end{figure}

Various NN-based control approaches have been investigated with guarantees of constraint satisfaction and recursive feasibility. An NN-based control method using the differential programming scheme with probabilistic guarantees on stability and constraint satisfaction was developed to stabilize PVTOL aircraft considering linear state and control constraints in~\citep{mukherjee2022neural}. Additionally, in~\citep{chen2022large}, the primal active set QP method employs the output of the NN-based policy as an initial guess for a warm starting and primal feasibility check, which ensures recursive feasibility and asymptotic stability. In~\citep{chen2018approximating} and~\citep{karg2020efficient}, the QP solver is used to compute the projected control inputs, which are the closest control inputs with respect to the NN prediction, which satisfy state and control constraints. In this paper, we consider the implementation of Time Shift Governor (TSG) using neural networks, see Figure~\ref{fig:blockDiagram}.

The TSG is an add-on parameter governor~\citep{kolmanovsky2006parameter} scheme that enforces state and control constraints for the closed-loop system guided by time-shifting the reference trajectory to which the closed-loop system responds, which is the target orbit in spacecraft applications. The model is used to predict the future constraint violations with the assumed time-shift parameter, and the smallest in magnitude value of the time shift, which is feasible under constraints, is chosen by solving a constrained optimization problem.
For constrained spacecraft formation flying problems along circular orbits, TSG was developed in~\citep{2016gregory} based on the CWH equations~\citep{clohessy1960terminal}. TSG was also developed for constrained spacecraft Rendezvous and Proximity Operations (RPO) on highly elliptic Earth orbits in~\citep{kim2024timeAIAA} and on halo orbits in cislunar setting in~\citep{kim2024timeACC}. Notably, in~\citep{kim2024timeACC,kim2024timeAIAA}, time shift optimization was performed with nonlinear models used for the prediction of constraint violation. Despite the TSG optimization problem being low dimensional, the computational time is increased due to the use of nonlinear and more complex models describing the spacecraft dynamics.

The Kolmogorov-Arnold Networks (KANs)~\citep{liu2024kan}, which are based on the Kolmogorov-Arnold representation theorem~\citep{kolmogorov1957representation}, have emerged as promising alternatives to classical MLP-based NNs. KAN demonstrates several advantages over MLP, including comparable or better accuracy of KANs with smaller model sizes in function fitting tasks, enhanced neural scaling laws, and improved interpretability~\citep{liu2024kan}. With the growing popularity of KAN, many variants specific to the learning tasks have been investigated~\citep{somvanshi2024survey}. For instance, KAN was developed to solve OCP for a two-dimensional heat equation, outperforming conventional MLPs in accuracy and efficiency~\citep{aghaei2024kantrol}. Compared to MLP, KAN was also employed for function approximation in online reinforcement learning (RL) and demonstrated comparable performance using fewer parameters~\citep{kich2024kolmogorov}. 
Despite the emerging interest and use of KAN in various domains, KAN remains unexplored in constrained control, including for space missions.

In this paper, we consider the use of KAN to address constrained control problems for space missions. We demonstrate the effectiveness of the approach in spacecraft rendezvous simulations with various constraints. We also illustrate the robustness of the approach through simulations of maneuvers due to multiple initial conditions, and show that KAN approximations have lower complexity in terms of the number of parameters, as compared to deep learning MLP approaches, while achieving accurate prediction in terms of validation loss with comparable computation time.
 
The remainder of this article is organized as follows. 
Section \ref{sec:tsg} presents the problem formulation and introduces the TSG algorithm and conditions to approximate the TSG function.
Section \ref{sec:cinn} formulates the spacecraft RPO problems in Earth orbit and introduces the TSG development for constraint enforcement with given nominal controllers. 
Section \ref{sec:rpo} presents the simulation results of the RPOs. 
Finally, Section \ref{sec:conclusion} concludes with a summary.

%% file: sections/tsg.tex
We consider a continuous-time dynamical system represented by $\dot{{x}}=f({x},{u}),$ where ${x} \in \mathbb{R}^{n_x}$ and ${u} \in \mathbb{R}^{n_u}$ are the system state and control input, respectively. We assume that the function $f$ is continuous over $\mathbb{R}^{n_x}\times \mathbb{R}^{n_u}$. The solution at time $t$, assuming the initial state is ${x}_0$, is denoted by $\phi({x}_0, {u}(\cdot), t)$. The system must adhere to state and input constraints, represented by ${x} \in \mathbb{X}$ and ${u}\in \mathbb{U}$. We let ${\alpha}({x},{x}_v )$ denote a nominal feedback law such that the closed-loop system with ${u}={\alpha}({x},{x}_v)$ is forward complete with unique trajectories and asymptotically stable at the target state ${x}_v$.

In~\citep{kim2024timeACC,kim2024timeAIAA}, the Time Shift Governor (TSG) is proposed for spacecraft orbital rendezvous problems in which the rendezvous target trajectory corresponds to the unforced model, ${x}_{c}(t)=\phi({x}_c(0),{0},t)$, and the mechanism for enforcing constraints is based on time-shifting ${x}_c$, i.e., replacing ${x}_v(t)=x_c(t)$ with ${x}_v(t) = {x}_c (t+t_{\rm shift})$ in the feedback law so that the closed-loop dynamics become $\dot{{x}}(t)=f({x}(t),{\alpha}({x}(t),{x}_c(t+t_{\rm shift})))$. Note that the state constraint $\mathbb{X}$ depends on ${x}_c$ in the spacecraft rendezvous problems, i.e., $\mathbb{X}=\mathbb{X}({x}_c)$, introducing further challenges. The time shift is determined by solving the following optimization problem at the time instant $t$,
\begin{align}
    & \min_{t_{\rm shift}\in \mathcal{T} } |t_{\rm shift}| \label{eq:opt_prob1}\\ 
    \text{subject to}&\quad {x}_{k+1}=\phi({x}_k,{u}_k,T_s)\in \mathbb{X},\; k=0,\cdots,N_p -1,\; {x}_0={x}(t), \nonumber\\ 
    &\quad {u}_{k}={\alpha}({x}_k, {x}_{v,k})\in \mathbb{U},\; k=0,\cdots,N_p -1, \nonumber\\ 
    &\quad {x}_{v,k+1}=\phi({x}_{v,k},{0},T_s),\; k=0,\cdots,N_p -2, \nonumber\\ 
    &\quad {x}_{v,0}=\phi({x}_c (t), {0}, t_{\rm shift}), \nonumber
\end{align}
where $\mathcal{T}$ is a set of time shifts over which the optimization is performed. $T_s$ is the time discretization period chosen so that the problem has a finite number of constraints and $N_p$ is a sufficiently long prediction horizon. The time shift determined by solving the optimization problem~\eqref{eq:opt_prob1} at time instant $t$ is applied over the time interval $[t,t+P_{\tt ref})$ where $P_{\tt ref} \geq T_s$ and then recomputed for given ${x}(t+P_{\tt ref})$ and ${x}_c (t+P_{\tt ref})$.

We note that in the actual implementation of TSG for spacecraft rendezvous, the state constraints are more complicated, e.g., they can depend on ${x}_c$. Additionally, terminal set constraints may also be imposed to maintain the recursive feasibility of $t_{\rm shift}$, see ~\citep{kim2024timeAIAA} for details. Thus, the optimal time shift parameter is defined by mapping
\begin{equation} \label{eq:true_tsg}
    t_{\rm shift}^{\ast}(t) = \pi^{\ast}({x}(t);{x}_c (t)).
\end{equation}
In the sequel, we consider offline approximation of $\pi^{\ast}$ using machine learning techniques. Let $\pi_{\theta}\approx \pi^{\ast}$, derived off-line, be such an approximation where $\theta$ denotes parameters of the approximating function, such as neural network weights. Suppose the values of $\pi^{\ast}$ are known at a finite number, $\eta$, of data points,
\begin{equation}
    \tilde{t}_{\rm shift}^{j} = \pi^{\ast}(\tilde{{x}}^{j},\tilde{{x}}_c^j),\; j=1,\cdots,\eta,
\end{equation}
and denote the set of sample states as
\begin{equation}
    \mathcal{X}_{\eta}=\{(\tilde{{x}}^{j},\tilde{{x}}_c^j), j=1,\cdots,\eta \}.
\end{equation}
Notably, properties of approximate explicit MPC feedback laws when the number of sample points increases have been studied in~\citep{canale2010set,canale2009set}, and conditions for closed-loop stability have been obtained. In the TSG setting, the closed-loop stability is maintained by the nominal feedback law, however, the use of approximations can degrade constraint enforcement, recursive feasibility, and time shift convergence properties. To improve the robustness of the TSG scheme against approximation errors, the constraints in the optimization problem~\eqref{eq:opt_prob1} can be imposed for all values of the time shift in the interval $[t_{\rm shift}-\zeta,t_{\rm shift}+\zeta]$ where $\zeta \geq 0$ is the upper bound on the expected approximation error. In this article, we focus on demonstrating the potential of recent advances in machine learning using CIKAN to approximate $\pi^{\ast}$ in spacecraft orbital rendezvous problems while leaving the study of conditions on the approximation that ensure TSG convergence and feasibility properties to future research.

%% file: sections/cinn.tex
Constraint-informed Neural Networks (CINNs) are neural networks (NNs) that are trained to solve supervised learning tasks while accounting for the imposed constraints. The loss function of CINN penalizes prediction errors and constraint violations, so it has the form, 
\begin{equation}
    \mathcal{L}_{\rm total} = \mathcal{L}_{\rm regression} + \mathcal{L}_{\rm CINN}.
\end{equation}
Using supervised learning, NNs can approximate a nominal control law that generates control inputs satisfying constraints. 

TSG generates an adjusted reference trajectory that depends on a single parameter, which is the time shift, to enforce state and control constraints. The optimal time shift, obtained with the nominal TSG approach, is the closest time shift value to zero, enforcing constraints. The TSG optimization problem is low dimensional as only a scalar time shift parameter is being determined, and, assuming a sufficiently long prediction horizon or the use of the terminal constraints, the time shift computed at the previous time instant remains a feasible solution at the current time instant.

\subsection{CINN Architecture}
We consider the use of CINN to approximate the optimal time shift mapping of TSG, $\pi^{\ast}$. To train the CINN, our initial choice of the loss function $\mathcal{L}_{\rm total}(\theta)$ combines the Mean Squared Error (MSE) $\mathcal{L}_{\rm regression}$ and the Mean Squared ReLU (MSReLU) $\mathcal{L}_{\rm CINN}$ terms:
\begin{equation} \label{eq:aug_msrelu}
    \mathcal{L}(\theta) = \mathbb{E}\left[( t^{\ast}_{\rm shift} - \pi_{\theta}(x))^2\right] + \theta_{c}\; \mathbb{E}\left[{\rm ReLU}(|t^{\ast}_{\rm shift}| - \pi_{\theta}(x))^2\right],
\end{equation}
where $\mathbb{E}$ denotes the expectation over all possible states encountered during training within a batch and $\theta_{\rm CINN}$ denotes a scalar weight of constraint violation. Note that we penalize smaller in magnitude predictions of $t_{\rm shift}^{\ast}$ as such smaller in magnitude values of the time shift may not enforce constraints over the prediction horizon. By minimizing this loss function during training CINN weights, $\theta$, are determined, i.e., the training objective of the CINN is to compute $\theta^{\ast}=\arg\min_{\theta\in \Theta} \mathcal{L}(\theta)$. 

Our initial numerical experiments using~\eqref{eq:aug_msrelu} revealed that CINN with the loss function in~\eqref{eq:aug_msrelu} struggles with learning small values of the time shift. As in the rendezvous missions with TSG the deputy spacecraft approaches the chief spacecraft either from the forward in track direction or from the backward in track direction, either $t_{\tt shift}^*(t)$ is nonnegative or nonpositive at all times $t$ for a specific maneuver. Therefore, our approach involves developing two separate approximations for the non-negative time shift and non-positive time shifts.  In this paper, we focus on the case of non-positive time shifts suitable for rendezvous from forward in track positions; for these, we developed an approximation $\tilde{\pi}_{\theta}(x)$ for $log(|t_{\tt shift}^*|)$ by minimizing a reformulated loss function with the logarithmic loss,
\begin{equation} \label{eq:log_aug_msrelu}
    \mathcal{L} (\theta) = \mathbb{E}\left[( \log ( |t^{\ast}_{\rm shift}| ) - \tilde{\pi}_{\theta}(x))^2\right] + \theta_{c}\; \mathbb{E}\left[{\rm ReLU}( \log ( |t^{\ast}_{\rm shift}| ) - \tilde{\pi}_{\theta}(x))^2\right],
\end{equation}
where $\mathcal{L}_{\rm regression}=\mathbb{E}\left[( \log ( |t^{\ast}_{\rm shift}| ) - \tilde{\pi}_{\theta}(x))^2\right]$ and $\mathcal{L}_{\rm CINN}=\theta_{c}\; \mathbb{E}\left[{\rm ReLU}( \log ( |t^{\ast}_{\rm shift}| ) - \tilde{\pi}_{\theta}(x))^2\right]$. Once trained, the output of the CINN is transformed to the time shift as
\begin{equation} \label{eq:out_t_shift}
    t_{\rm shift} = \pi_{\theta^{\ast}}(x) = \exp \left(\tilde{\pi}_{\theta^{\ast}} (x)\right),
\end{equation}
A similar solution can be developed by approximating the nonnegative time shifts for rendezvous maneuvers from backward in track.

Given an initial time shift and the set $\mathcal{T}$, the time shift can be updated by the CINN-based TSG algorithm, described in Algorithm~\ref{alg:hybrid_model}.
\begin{algorithm} 
\DontPrintSemicolon
\caption{CINN-based TSG algorithm}
\label{alg:hybrid_model}
\begin{algorithmic}[1]
\State \textbf{Input} ${x}_k={x}(t_k) \in \mathbb{X},\; \hat{t}_{{\rm shift},k-1}=\hat{t}_{\rm shift}({x}_{k-1})\in \mathcal{T}_{k-1},\; {x}_c = {x}_c(t_k),\; {u}(\cdot)$
\State \textbf{Output} $\hat{t}_{{\rm shift},k},\; \mathcal{T}_k $
\State Predict a time shift candidate $t_{\rm shift}^{\tt cand}\gets \pi_\theta ({x})$
\State \textbf{if} $t_{\rm shift}^{\tt cand} \in \mathcal{T}_{k-1}$
\State $\quad$ \textbf{if} $t_{\rm shift}^{\tt cand}$ ensures constraint satisfaction over a prediction horizon.
\State $\quad$ $\quad$ $\hat{t}_{{\rm shift},k}\gets t_{\rm shift}^{\tt cand}$
\State $\quad$ \textbf{else}
\State $\quad$ $\quad$ $\mathcal{T}\gets \{ t_{\rm shift}\in \mathbb{R}: \hat{t}_{{\rm shift},k-1} \leq t_{\rm shift} \leq t_{\rm shift}^{\tt cand} \}$
\State $\quad$ $\quad$ $\hat{t}_{{\rm shift},k}\gets \pi^{\ast} ({x}_k;{x}_c, {u}(\cdot), \mathcal{T})$ 
\State \textbf{else}
\State $\quad$ $\hat{t}_{{\rm shift},k}\gets \pi^{\ast} ({x}_k;{x}_c, {u}(\cdot), \mathcal{T}_{k-1})$ 
\State \textbf{end if}
\State $\mathcal{T}_k \gets \{ t_{\rm shift}\in \mathbb{R}: \hat{t}_{{\rm shift},k} \leq t_{\rm shift} \leq 0 \}$
\end{algorithmic}
\end{algorithm}
 
\subsection{Constrained-informed Kolmogorov Arnold Networks}
The KAN employs the learnable basis functions ``on edges'' and sums on a node without any nonlinearities. An activation function in the original KAN architecture~\citep{liu2024kan} is of the form
\begin{equation} \label{eq:residual_act_fxn}
    \tilde{\phi}(x) =\tilde{\phi}(x ;{\theta}, {\beta}, {\alpha}) = {\beta} \cdot b(x) + \alpha \cdot {\rm spline}(x; {\theta}),
\end{equation}
where $x$ is an input, ${\rm spline}(x; \theta)=\sum_i \theta_{i} B_i (x)$ denotes the spline function with learnable coefficients ${\theta}$, $b(x) = {\rm SiLU}(x) = x/(1 + e^{-x})$, and $\alpha$ and $\beta$ denote learnable coefficients. Spline functions are piecewise polynomial functions with high expressiveness and can approximate any continuous function. KAN can approximate various complex nonlinear functions by adjusting the parameters and coefficients of the activation function.

After the original KAN was introduced, other variants of KAN have been proposed. The Gaussian Radial Basis Function-based KAN (GRBF-KAN)~\citep{li2024kolmogorovarnold} uses
\begin{equation} \label{eq:residual_act_grbf}
    \tilde{\phi}(x)= \sum_{i} \theta_{i} \exp \left( -\frac{\| x-c_{i} \|}{ 2 h^2} \right),
\end{equation}
where $\theta_i$ denotes learnable coefficients, $h$ is a width parameter, and $c_i$ are the center points. The Reflectional SWitch Activation Function-based KAN (RSWAF-KAN)~\citep{Athanasios2024} uses 
\begin{equation} \label{eq:residual_act_fxn2}
    \tilde{\phi}(x)= \sum_{i} \theta_{i} \left( 1-\left(\tanh \left(\frac{x-c_{i}}{h} \right) \right)^2 \right).
\end{equation}

Using the activation functions, the activation value associated with the $(l+1,j)$ neuron can be written as
\begin{equation} \label{eq:node_sum}
    x_{l+1,j}=\sum_{i=1}^{n_l} \tilde{\phi}_{l,j,i}(x_{l,i}),\; j=1,\cdots,n_{l+1},
\end{equation}
where $\tilde{\phi}_{l,j,i}$ denotes the activation function that connects $(l,i)$-neuron and $(l+1,j)$-neuron, $l$ is the index of layers, and $i,j$ is the index of neuron in the $l$-th layer. Then,~\eqref{eq:node_sum} can be expressed in matrix form as
\begin{equation} \label{eq:node_sum_mat}
    {x}_{l+1}=\begin{bmatrix}
        \tilde{\phi}_{l,1,1}(\cdot)&\tilde{\phi}_{l,1,2}(\cdot)& \cdots & \tilde{\phi}_{l,1,n_l}(\cdot)\\
        \tilde{\phi}_{l,2,1}(\cdot)&\tilde{\phi}_{l,2,2}(\cdot)& \cdots & \tilde{\phi}_{l,2,n_l}(\cdot)\\
        \vdots&\vdots&  & \vdots\\
        \tilde{\phi}_{l,n_{l+1},1}(\cdot)&\tilde{\phi}_{l,n_{l+1},2}(\cdot)& \cdots & \tilde{\phi}_{l,n_{l+1},n_l}(\cdot)\\
    \end{bmatrix}{x}_l = \Phi_l {x}_l,
\end{equation}
where $\Phi_l $ denotes the matrix with function elements corresponding to the $l$-th KAN layer and matrix-vector multiplication in~\eqref{eq:node_sum_mat} is understood in a sense of applying functions in the matrix to components of the vector $x_l$. A general KAN network, ${\rm KAN}:\mathbb{R}^{n_0}\to \mathbb{R}$, with $L$ layers can be represented as
\begin{equation} \label{eq:kan_layers}
    {\rm KAN}({x}_{\rm in})= (\Phi_{L-1} \circ \Phi_{L-2} \circ \cdots \circ \Phi_{1} \circ \Phi_{0} ) {x}_{\rm in},
\end{equation}
where ${x}_{\rm in}\in \mathbb{R}^{n_{0}}$ is an input vector. In this work, we train a KAN-based model, $\pi_\theta$, in the form~\eqref{eq:kan_layers} to approximate $\pi^{\ast}$ in~\eqref{eq:true_tsg}. Reference \citep{liu2024kan} discusses the approximation theory that pertains to the convergence of the approximations as the number of points in the sample data set grows.

%% file: sections/spacecraft.tex
\begin{figure} [!]
     \centering     
     \includegraphics[width=0.51\linewidth]{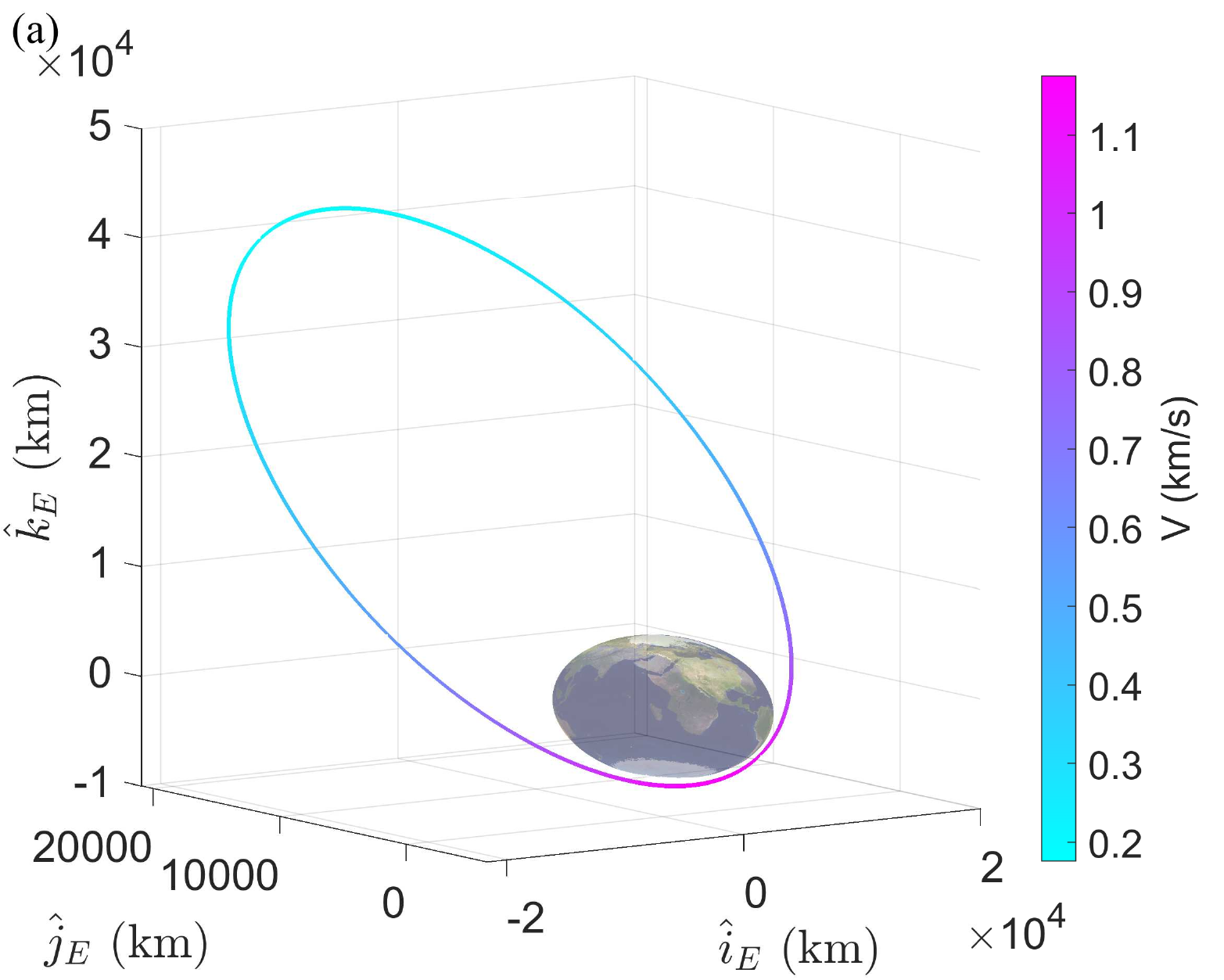}\;
     \includegraphics[width=0.47\linewidth]{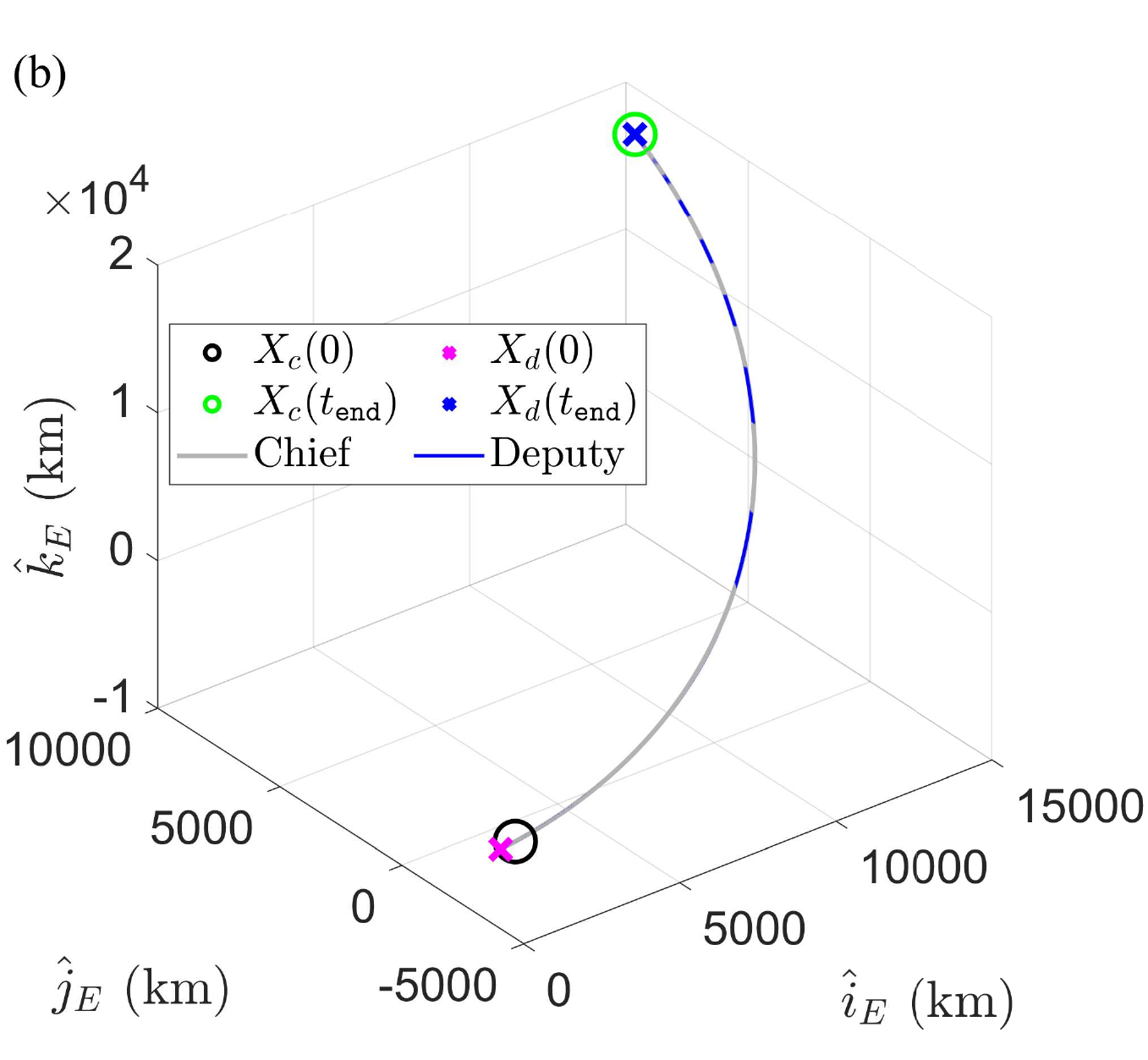}
     \caption{(a) The reference highly elliptic orbit, and (b) a sample RPO mission, expressed in the inertial frame.}
     \label{fig:ref_traj} \vspace{-0.2 in}
\end{figure}

Spacecraft Rendezvous and Proximity Operations (RPO) missions involve two spacecraft: a secondary spacecraft (Deputy), denoted as a subscript $d$, approaching a primary spacecraft (Chief), denoted as a subscript $c$, while adhering to mission-specific constraints. In this work, we demonstrate the application of Constrained-informed Kolmogorov Arnold Networks (CIKAN) to RPO problems in a highly elliptic orbit characterized by the classical orbital elements in Table~\ref{tab:coe_ref}, illustrated in Figure~\ref{fig:ref_traj}(a). Note that $a, e, i, \Omega, \omega, \nu,$ and $ T_{\rm period}$ denote the semi-major axis, eccentricity, inclination, right ascension of the ascending node, argument of periapsis, true anomaly, and one orbital period, respectively. RPO missions in the highly elliptic orbit introduce significant challenges: 1) the relative motion dynamics vary substantially in different parts of the orbit and 2) linearized dynamics models, e.g., Tschauner-Hempel equations~\citep{tschauner1965rendezvous}, are not able to accurately represent the relative dynamics between widely separated spacecraft. However, elliptic orbits are the second most common orbit type; in particular, they provide enhanced observation capabilities and increased coverage duration over high-latitude regions.

The spacecraft dynamics are described in the setting of the Two-Body Problem, i.e., $$\ddot{{r}}_i=-\frac{\mu}{\|r_i\|^{3}}{r}_i  + {u}_i, i\in \{c,d\},$$ where ${r}=p({x})\in \mathbb{R}^3$ is the spacecraft position vector, ${x}=[p^{\sf T}({x})\; v^{\sf T}({x})]^{\sf T}\in \mathbb{R}^6$ is the spacecraft position and velocity vector, $p:\mathbb{R}^6\to\mathbb{R}^{3},v:\mathbb{R}^6\to\mathbb{R}^{3}$ denote functions which represent position and velocity, respectively. 
\begin{table}[htbp] 
\caption{Classical orbital elements and orbital period.}
\begin{center}
\begin{tabular}{ccccccc}
\hline \hline
$a$ (km) & $e$ (-)& $i$ (deg)& $\Omega$ (deg)& $\omega$ (deg)& $\nu$ (deg) & (hrs) \\
\hline
26646.6808 & 0.74& 62.8& 0.0& 280 & 0 & 12.0247 \\
\hline \hline
\end{tabular}
\label{tab:coe_ref}
\end{center} \vspace{-0.2 in}
\end{table}

As in \citep{kim2024timeAIAA}, we employ the Frozen-in-time Riccati Equation (FTRE)-based controller as the nominal controller for the Deputy spacecraft; this controller is based on the solution to an unconstrained infinite-time horizon Linear Quadratic OCP for the linearized dynamics corresponding to the linearization at $x_v(t)$. To simplify the implementation, the FTRE gains are pre-computed offline 
for different values of the true anomaly along the nominal orbit of the Chief spacecraft and stored. If the virtual target adjusted by TSG is at $x_v(t)$ at time $t$, then the corresponding value of the FTRE-based controller gain is computed through interpolation. This process provides time-varying gains, which are also assumed in TSG prediction.


We consider three constraints in the RPO simulations. First, the Deputy spacecraft must operate within a Line of Sight (LoS) characterized by a half-cone angle, $\alpha_{\tt LoS}=20^{\circ}$,  with respect to the docking port during RPO. Assuming the docking port points the opposite to the Chief spacecraft's velocity vector, the LoS constraint is time-varying when expressed in the inertial frame while it is time-invariant in a local frame, e.g., Velocity-Normal-Binormal (VNB) frame centered at the Chief spacecraft. Second, due to the physical limitations of thruster(s), control input acceleration is limited by the maximum thrust value, $u_{\tt max}=0.5$ m/s$^2$. To handle the thrust limit, saturation is used on the output of the nominal controller to limit the control input magnitude to $u_{\tt max}$ while preserving its direction. When TSG performs the prediction, it accounts for this saturation in the closed-loop model used for forward propagation. Lastly, we consider the bound on the approach velocity as a function of approach distance, which is defined by a linear decrease with a slope, $\gamma_2 =20$ rad/s, and an offset, $\gamma_3=10^{-3}$ km/s; this constraint is only activated when the Deputy spacecraft is within the distance $\gamma_1=5$ km from the Chief spacecraft. Therefore, the sets where state and control constraints hold have the form, $\mathbb{U}=\{{u}\in\mathbb{R}^3: h_2 = \| {u}\| - u_{\tt max}\leq 0 \}, \mathbb{X}=\{{x}_{d,c} \in \mathbb{R}^{12}: {x}_{d,c}=({x}_d, {x}_c) \in h_j({x}_{d,c}), j\in \{1,3\} \}$, where $h_1=-v(x_c)^{\sf T} p(x_d-x_c) + \cos(\alpha_{\tt LoS}) \|v(x_c)\| \|p(x_d - x_c)\| \leq 0$ and $ h_3=\| v(X_d - X_c)\|-\gamma_1 \|p(X_d -X_c)\|-\gamma_2\leq 0$, if $\|p(x_d - x_c)\|\leq \gamma_1$.


\begin{table}[t!] \vspace{-0.2 in}
\caption{Comparison of model performances with specifications}
\begin{center}
\begin{tabular}{cccccc}
\hline \hline
Model & Model complexity & RMSE (s)  & \multicolumn{2}{c}{Computation time (s)}& Delta V \\
($n$, $\ell$, $G$)&$(\mathcal{L}_{\rm train},\mathcal{L}_{\rm val})\times 10^{-3}$ & & average & worst case& average (km/s)\\
\hline
TSG& -   & - & 0.0871 & 0.8546 & 3.8741 \\
\hline
CIKAN & 1,638,400 & \multirow{2}{*}{0.9652} & \multirow{2}{*}{0.0329}   & \multirow{2}{*}{0.5590}   & \multirow{2}{*}{3.4965}\\
 (128,5,17)& (1.3218, 1.2615) &  &  &    & \\
\hline
GRBF-CIKAN& 122,880 & \multirow{2}{*}{0.8220} & \multirow{2}{*}{0.0451}   & \multirow{2}{*}{1.0339 }   & \multirow{2}{*}{3.5489}\\
(32,6,17)& (1.4758, 1.1540) &  &  &    & \\
\hline
RSWAF-CIKAN & 491,520 & \multirow{2}{*}{1.2906} & \multirow{2}{*}{0.0379}   & \multirow{2}{*}{0.5435}   & \multirow{2}{*}{3.4807}\\
  (64,6,17)& (1.5729, 1.2610) &  &  &    & \\
\hline
CINN-ReLU & 209,715,208 & \multirow{2}{*}{0.7483} & \multirow{2}{*}{0.0543}   & \multirow{2}{*}{1.3284}   & \multirow{2}{*}{3.6593}\\
  (2048,5,-)& (2.7907, 1.6458) &  &  &    & \\
\hline
CINN-GELU & 5,242,880 & \multirow{2}{*}{1.1202} & \multirow{2}{*}{0.0299}   & \multirow{2}{*}{0.5292}   & \multirow{2}{*}{3.4549}\\
  (1024,5,-)& (2.8983, 1.8538) &  &  &    & \\
\hline \hline
\end{tabular}
\label{tab:comparison}
\end{center} \vspace{-0.3 in}
\end{table}

To train the CIKAN and CINN, based on more conventional MLP architectures, we employ a data set, $\mathcal{D}=\{({x}_{d,c}^{(l)} ,t_{\rm shift}^{\ast})^{(l)} \}_{l=1}^{1{\rm M}}$, which consists of 1 million state-time shift pairs. 
The training is performed on the data with negative time shifts.
We optimize the hyperparameters of NN over 100 trials using the Tree-structured Parzen estimator (TPE)~\citep{bergstra2011algorithms}. We train $\{\theta \}$ of an NN using the Adam with decoupled weight decay (AdamW)~\citep{loshchilov2017decoupled} provided in the PyTorch library, the dataset, $\mathcal{D}$, to minimize the loss function, $\mathcal{L}(\theta)$, in~\eqref{eq:log_aug_msrelu}, and the exponential hyperbolic learning scheduler~\citep{kim2024hyperboliclr}, which is used in the updated process, i.e., $\theta^{(q+1)}_{i}=\theta^{(q)}_{i}-\Delta_{lr}\frac{\partial \mathcal{L}}{\partial \theta_{i}}$, where $\Delta_{lr}$ is the learning rate.

We conduct the simulations on Julia 1.11.1 platform using an Intel i9-12900HK 2.50 GHz processor running Windows 11 and 32.0 GB of RAM. Table~\ref{tab:comparison} presents model specifications and comparisons of performances, such as NN's loss values in~\eqref{eq:log_aug_msrelu} and model complexity $n_{\theta}$, and CINN-based TSG's root mean squared error (RMSE), computation time, and fuel consumption. As stated in~\citep{liu2024kan}, the model complexity represented by the number of parameters is defined as $n_{\theta} = n^2 \times \ell \times (G+k_d)$, where $\ell, n, G, k_d$ denote layers, nodes, grid size, and degree of the basis spline function, respectively. 

\begin{figure} [!]
     \centering
     \includegraphics[width=0.32\linewidth]{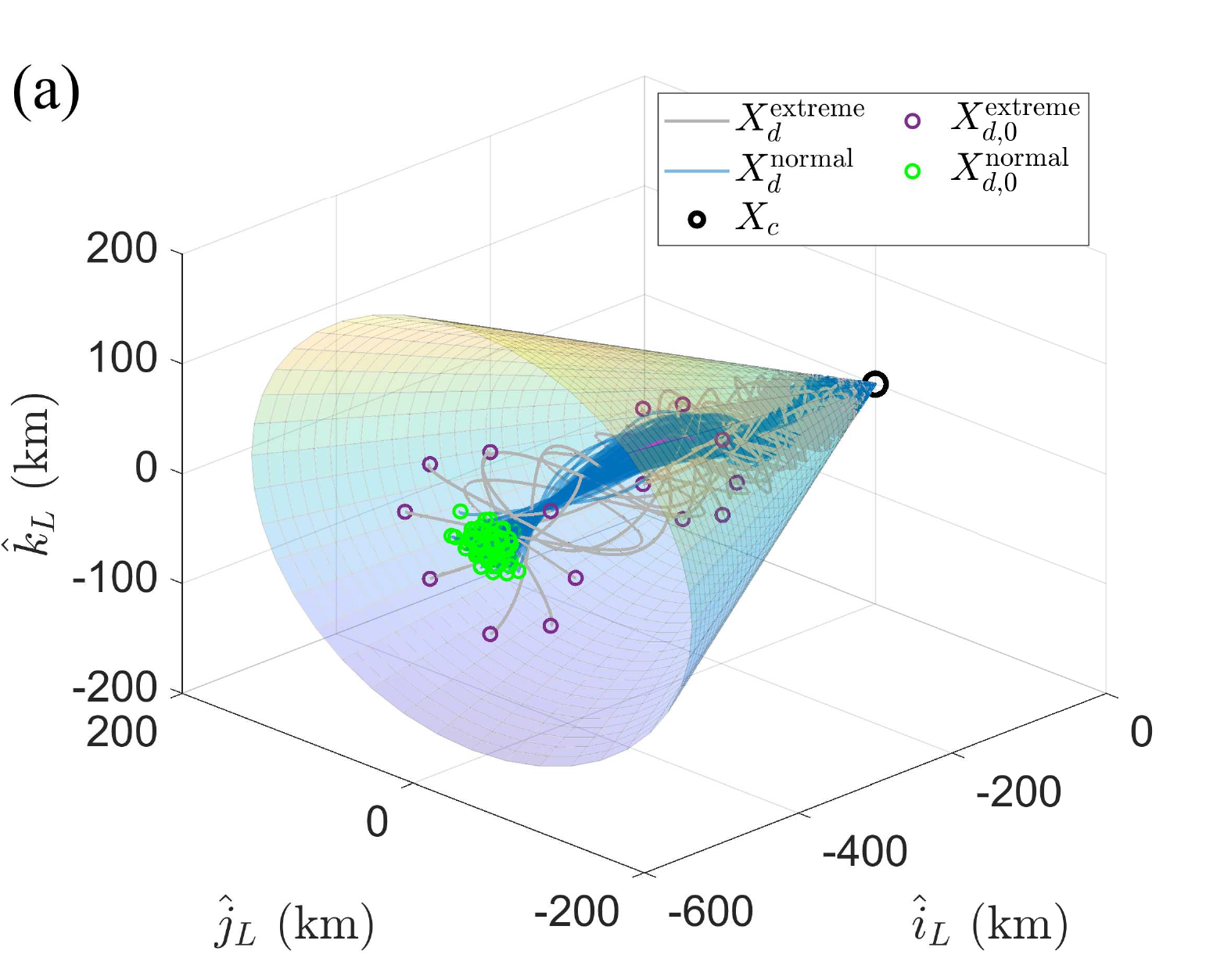}\;
     \includegraphics[width=0.32\linewidth]{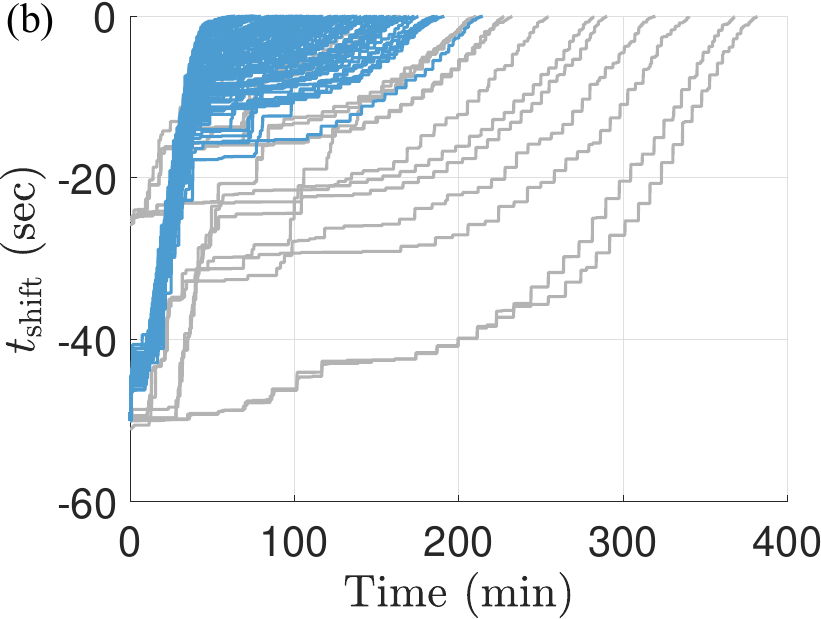}\;
     \includegraphics[width=0.31\linewidth]{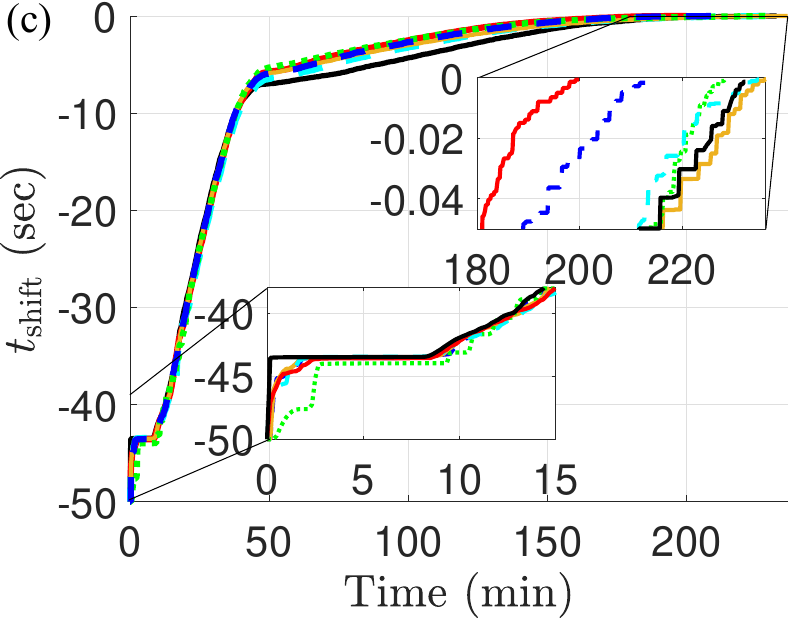}\\
     \includegraphics[width=0.32\linewidth]{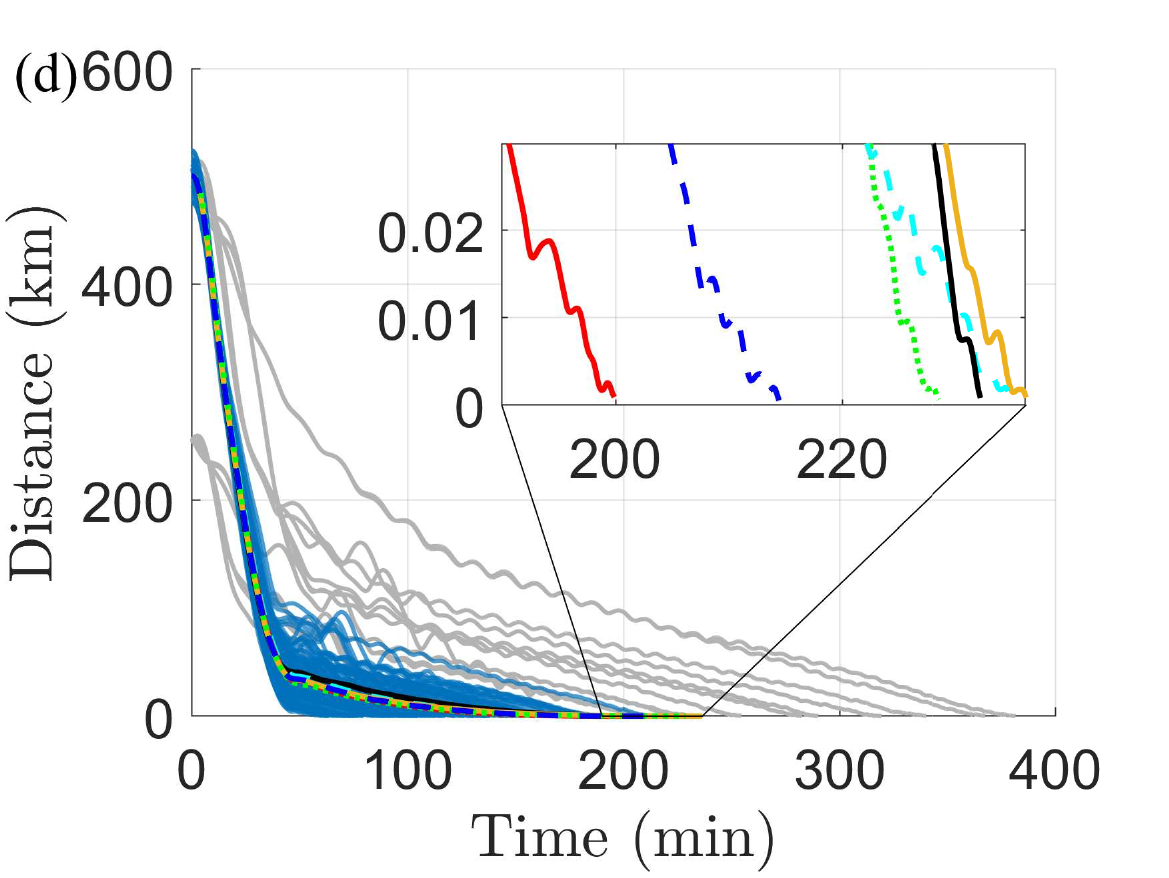}\; 
     \includegraphics[width=0.34\linewidth]{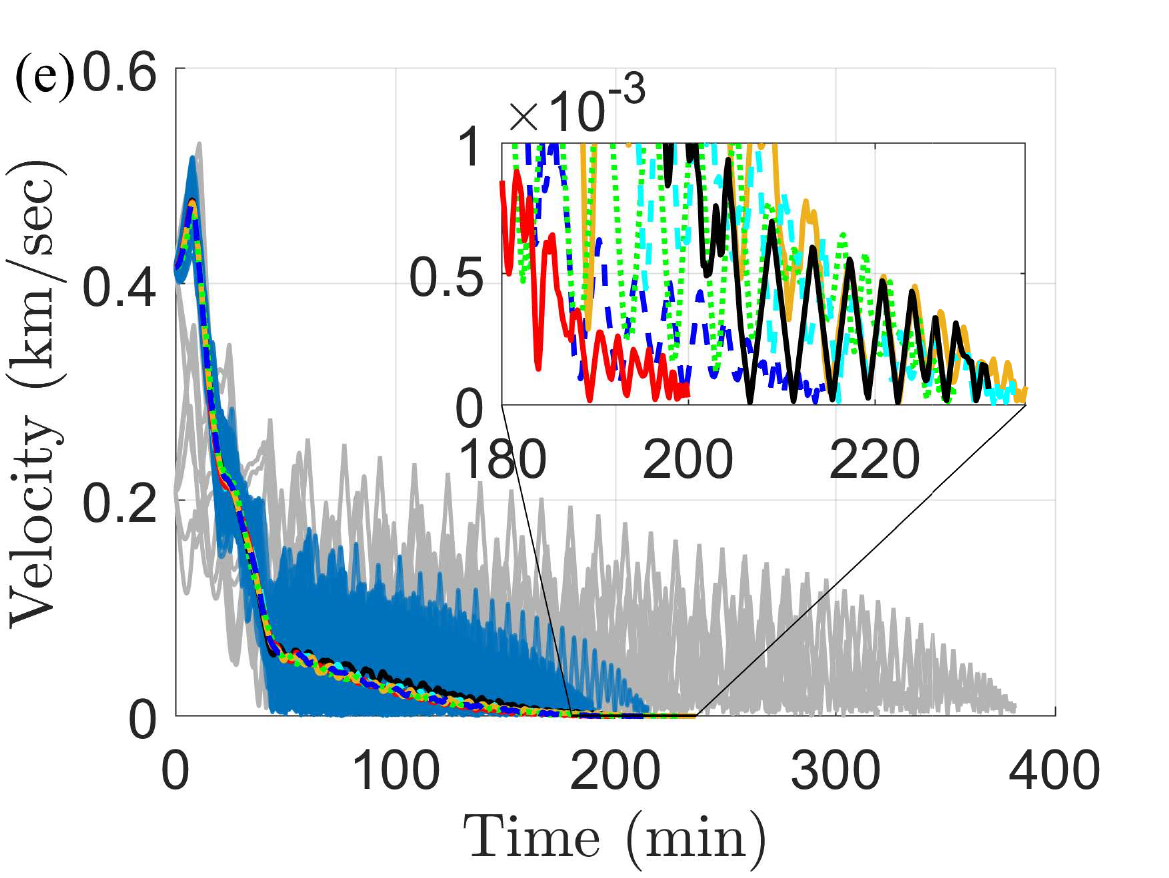}\;
     \includegraphics[width=0.21\linewidth]{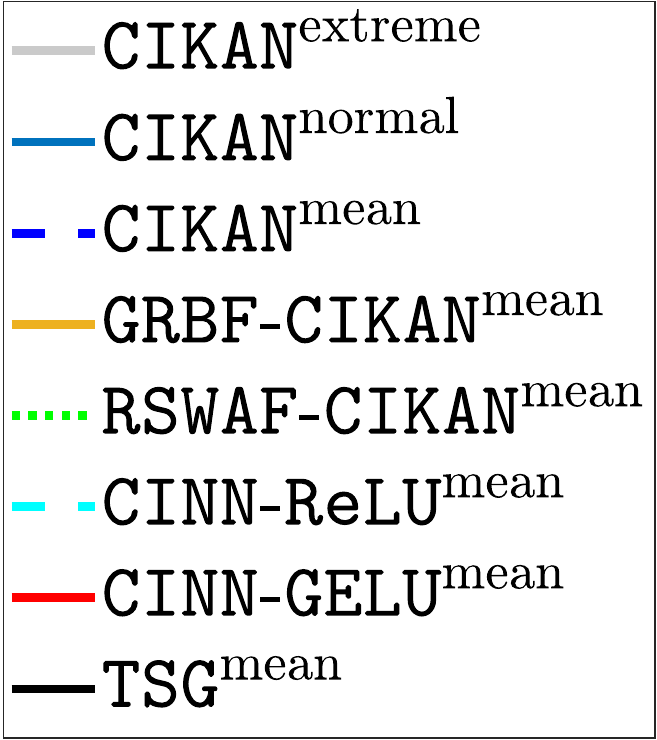}
     \caption{Simulations of extreme cases and Monte Carlo runs using various models: (a) Relative trajectories of Deputy spacecraft expressed in the VNB frame and Time histories of (b) time shift parameter $t_{\rm shift}$, (c) mean of time shift $\mathbb{E}[t_{\rm shift}]$, (d) relative distance, and (e) relative velocity of the Deputy spacecraft with respect to the Chief spacecraft.}
     \label{fig:sc_state} \vspace{-0.3 in}
\end{figure}
Figure~\ref{fig:sc_state} demonstrates the robustness of the six different TSG models, such as CIKAN (blue dashed), GRBF-CIKAN (yellow solid), RSWAF-CIKAN (green dotted), CINN-ReLU (cyan dashed), CINN-GELU (red solid), and conventional method (black solid), by presenting the results of extreme cases (gray) and the Monte Carlo simulations (blue) with normally distributed 100 initial states, $p({x}-{x}_c)\sim \mathcal{N}(0_{3\times 3}, {\rm diag}(10^2,10^2,10^2))$ in the unit of km. As an illustration, in Figure~\ref{fig:ref_traj}(b), which is expressed in an inertial frame, we consider one of the RPO missions starting at the periapsis of the reference orbit where the changes in the relative motion dynamics are most substantial. In Figure~\ref{fig:sc_state}(a), which is expressed in a local frame, the Deputy spacecraft starting from perturbed initial states satisfies the LoS cone constraint using the CIKAN-based TSG. Figure~\ref{fig:sc_state}(b) shows that the time shift parameter converges to zero in finite time for all scenarios, and, with the FTRE-based controller, they complete the spacecraft rendezvous mission is successfully completed by converging to the Chief spacecraft, as shown in Figure~\ref{fig:sc_state}(d),(e).

Starting with the normally distributed initial states, we present the time history of mean values using six different models to demonstrate the approximate model performance compared to the conventional TSG approach illustrated in Figure~\ref{fig:sc_state}(c),(d),(e). Comparing the five NN models, KAN models, such as CIKAN, GRBF-KAN, and RSWAF-CIKAN, show smaller training and validation loss than MLP models, such as CINN-ReLU and CINN-GELU, with smaller model complexities. Particularly, the GRBF-KAN shows the best performance in terms of validation loss with the smallest number of parameters. On the other hand, the CINN-ReLU shows the smallest RMSE value, while using the largest number of parameters and computation time on average. 
With the second largest number of parameters, the CINN-GELU achieves the simulation with the lowest computation time on average and in the worst case, while using the smallest Delta V, which is a metric for fuel consumption.

The NN-based models achieve advantages by replacing numerical computations with neural network predictions for time shift calculations. As shown in Figure~\ref{fig:sc_state}(c), all approximate models generate comparable time shift evolution patterns to the conventional TSG. These models introduce approximation errors when solving the optimization problem in~\eqref{eq:opt_prob1}, resulting in different state trajectories, as illustrated in Figure~\ref{fig:sc_state}(d) and (e), that lead to improved Delta V performance.

\begin{figure} 
     \centering
     \includegraphics[width=0.31\linewidth]{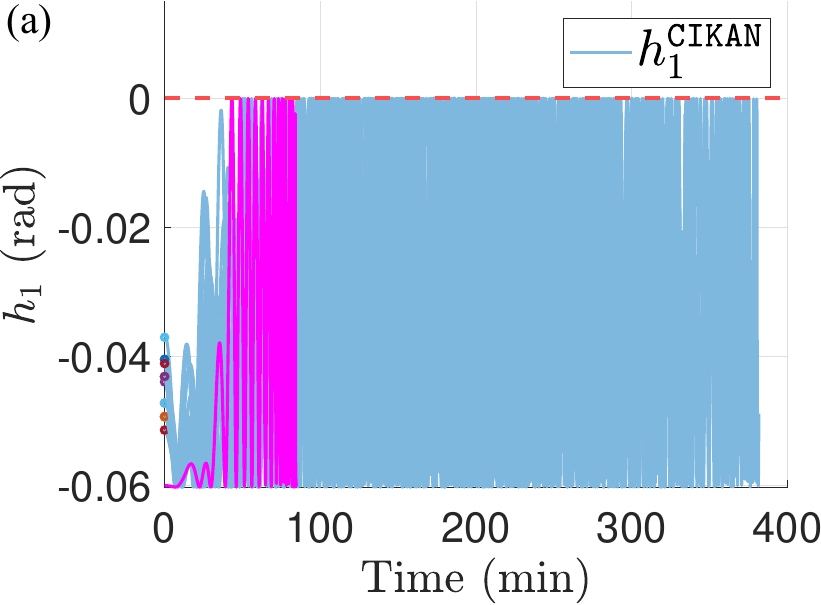}\; 
     \includegraphics[width=0.31\linewidth]{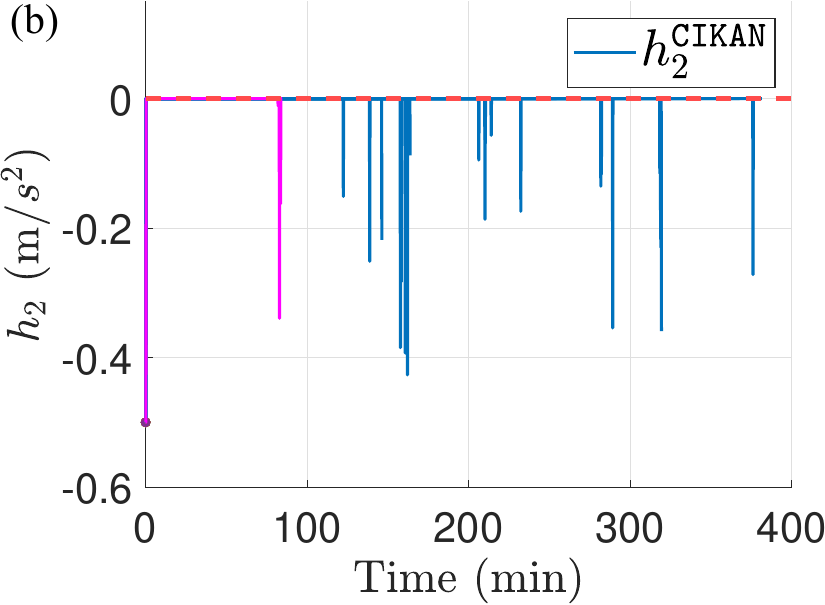}\; 
     \includegraphics[width=0.31\linewidth]{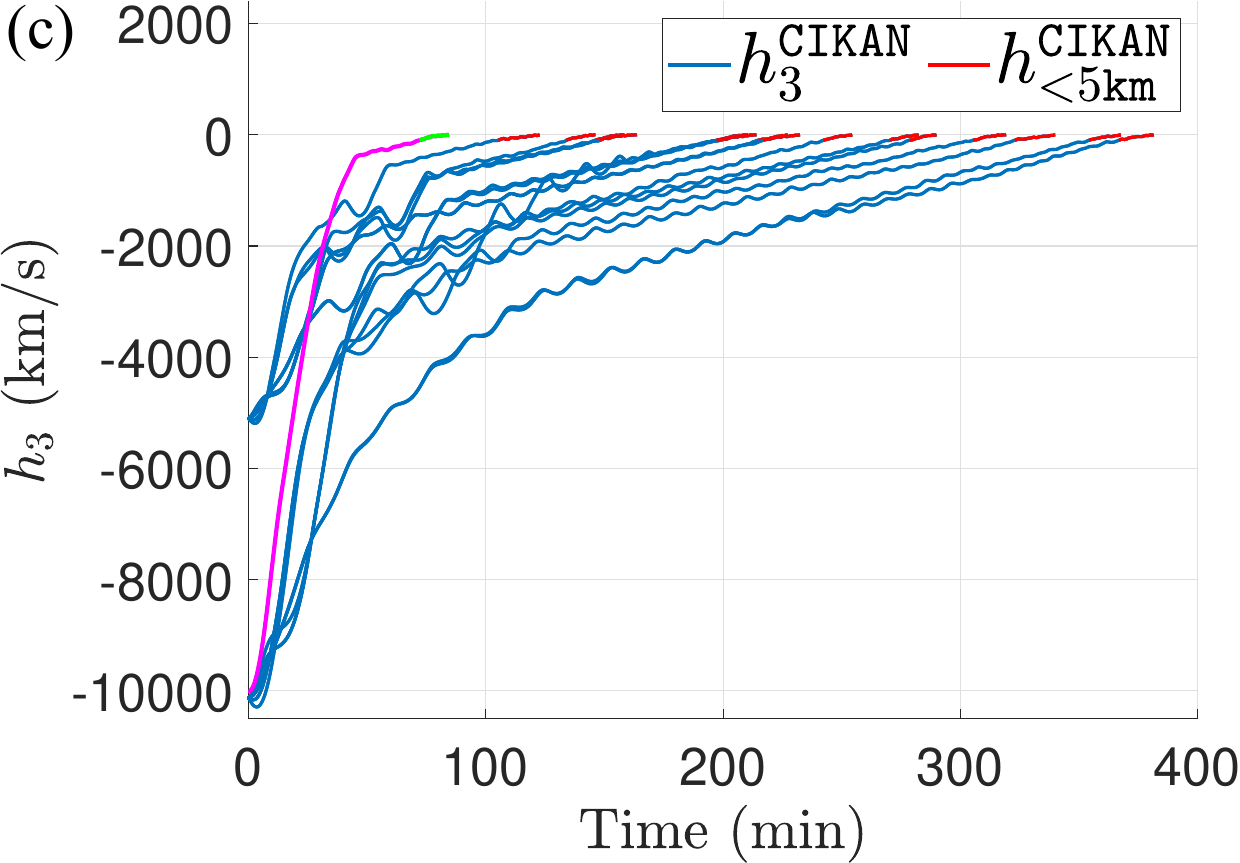}
    \caption{Time histories of (a) LoS cone constraints; (b) thrust magnitude limit; (c) approach velocity limit.}  
    \label{fig:const} \vspace{-0.3 in} 
\end{figure}

Figure~\ref{fig:const} presents the time history of inequality constraints (blue) using the CIKAN-based TSG starting from extreme initial states far from the Chief spacecraft's velocity direction for which feasible time shifts exist at the initial time instant. Notably, CIKAN-based TSG is capable of enforcing the LoS constraint, $h_1$, thrust magnitude limit, $h_2$, and approach velocity constraint, $h_3$. Although the results from the other three models (conventional TSG, CINN-based TSG, GRBF-CIKAN-based TSG, and RSWAF-CIKAN-based TSG) are not included, they also guide the Deputy spacecraft to the Chief spacecraft without constraint violations.



%% file: sections/conclusion.tex
This paper considered an opportunity in integrating machine learning and constrained control.  The use of Kolmogorov-Arnold (KAN) neural networks has shown promising results for approximating the solution mapping of a constrained optimization problem, which determines the value of the time shift parameter of the Time Shift Governor (TSG).   The TSG is an add-on scheme to the nominal closed-loop system that adjusts a time-shifted reference in order to enforce the constraints. The solution based on three different KAN implementations has been compared to two MLP-based solutions. The results show that these CINN-based models outperform the conventional TSG in average computation time and fuel consumption, while successfully accomplishing the constrained spacecraft rendezvous maneuver on an elliptic orbit. Compared to MLP-based CINNs, CIKANs have lower model complexity, smaller training and validation loss, and comparable simulation performance.